\documentclass[preprint,nofootinbib,amsmath]{revtex4}
\usepackage{graphicx}

\begin{document}

\title{Anisotropic Inflation and the  Origin of Four Large Dimensions}
\author{C.    Armend\'ariz-Pic\'on}  \email{armen@oddjob.uchicago.edu}
\affiliation{Enrico  Fermi Institute and  Department of  Astronomy and
  Astrophysics,    \\   University   of    Chicago.}    \author{Vikram
  Duvvuri}\email{duvvuri@theory.uchicago.edu}       \affiliation{Enrico
  Fermi Institute and Department of Physics, University of Chicago.}

\begin{abstract}
  In the context of $(4+d)$-dimensional general relativity, we propose
  an inflationary scenario wherein  $3$ spatial dimensions grow large,
  while $d$ extra  dimensions remain small. Our model  requires that a
  self- interacting $d$-form acquire  a vacuum expectation value along
  the extra dimensions. This causes $3$ spatial dimensions to inflate,
  whilst keeping the size of  the extra dimensions nearly constant. We
  do not require an additional stabilization mechanism for the radion,
  as  stable  solutions exist  for  flat,  and  for negatively  curved
  compact extra dimensions.   From a four-dimensional perspective, the
  radion does not couple to  the inflaton; and, the small amplitude of
  the  CMB   temperature  anisotropies  arises   from  an  exponential
  suppression of fluctuations, due to the higher-dimensional origin of
  the  inflaton.  The  mechanism triggering  the end  of  inflation is
  responsible,  both,  for  heating  the universe,  and  for  avoiding
  violations of the equivalence  principle due to coupling between the
  radion and matter.
\end{abstract}

\maketitle

\section{Introduction}
Most  theories which  attempt  a unification  of  gravity with  matter
require  the existence  of additional  spatial dimensions.  This gives
rise  to a  pressing  question:  Why do  the  extra dimensions  remain
unobserved? Traditionally, this question has been addressed by arguing
that  the extra  dimensions are  ``small'' compared  to experimentally
accessible  scales, and  hence, hitherto  undetected.  Insofar  as the
dynamics of these dimensions is gravitational, it is believed that the
relevant scale is the Planck length. Thus, it is natural for the extra
dimensions   to  be   Planck-sized.  However,   the  spirit   of  this
``naturalness''  argument  requires that  we  apply  it to  dimensions
containing the observable universe,  too. This leads to the conclusion
that we populate a Planck-sized universe, in obvious disagreement with
observations.
 
The latter of  the two problems stated above  was one, amongst others,
which inflation  was originally  intended to address  \cite{Guth}.  In
the  conventional  picture,  an  initially  Planck-sized  universe  is
``blown up'', by  a sufficiently long stage of  inflation, into a size
much larger  than that of  our present horizon, whilst  also producing
many other  features of  the universe we  observe today  \cite{Linde}. 
Hence, from  the inflationary  point of view,  the problem is  not the
``smallness'' of the extra  dimensions, but rather the ``largeness''of
the  observable ones.   In  other  words, we  must  explain how  three
spatial  dimensions  inflated  while  the remaining  ones  kept  their
natural, Planckian size.

Conventionally,  inflation is driven  by a  scalar field.   Instead of
considering theories where a  particular subset of dimensions has been
singled out, as in brane-world  models \cite{branes}, in this paper we
shall deal with theories where  all spatial dimensions are equivalent. 
Then,  since a  scalar  does not  single  out any  direction in  space
either, it  generally yields accelerated expansion in  all dimensions. 
In order  to preferentially inflate only four  dimensions, a mechanism
to keep  the extra  dimensions at constant  size is required.   From a
four-dimensional  perspective, the  size  of the  extra dimensions  is
characterized by a single scalar field, the radion.  Even if we manage
to  stabilize the  radion, say,  by the  addition of  a  magnetic flux
threading the  extra dimensions  \cite{fluxes}, it is  unclear whether
the extra dimensions would remain stable during an inflationary phase.
In fact, generally  the inflaton is expected to  couple to the radion,
and it is likely that this coupling does violence to the stabilization
mechanism during inflation \cite{Geraldine,MaPe}. The stabilization of
compactified    spaces    has     been    extensively    studied    in
\cite{GuntherZhuk}.

In this paper  we propose an alternative minimal  model which achieves
inflation  in  four dimensions  whilst  keeping  the extra  dimensions
``small''. In order  to facilitate this, we require  that inflation be
driven  by  a  form,  rather  than  a scalar  field.  The  benefit  of
introducing a form lies in the  fact that, unlike a scalar, a form may
point in several  directions. By wrapping the extra  directions with a
form,  we are able  to introduce  an anisotropy  that is  necessary to
distinguish four  dimension from the  additional ones. The  details of
four-dimensional inflation are then determined by the self-interaction
of the form. Furthermore, it is not necessary to provide an additional
stabilization mechanism for the radion, and the interactions of the 
inflaton allow it to decouple naturally from the radion.

Massless   forms  are   ubiquitous  in   superstring  theory   and  in
supergravity.  In string theory, forms are a particularly crucial part
of the  spectrum since  a $(p+1)$-form couples  to a  $Dp$-brane, much
like  a  vector  boson   ($1$-form)  couples  to  a  charged  particle
($0$-brane) in  gauge theories.  We,  however, will require  a massive
form.  While massive  forms  are  not as  familiar  as their  massless
counterparts,  they are  certainly not  unheard of.  For  instance, in
supergravity,  a  $2$-form  may   acquire  mass  by  eating  a  vector
\cite{Romans}.

Forms also have a history  in cosmology. Form fluxes in compact spaces
have long  been considered as  a way to stabilize  internal dimensions
\cite{fluxes}.  Okada  studied how the presence of  these fluxes might
bring  about  radion  induced  inflation \cite{Okada}.   The  role  of
$2$-forms in Pre-Big Bang scenarios \cite{GasperiniVeneziano} has been
discussed     by     several     authors     (see     for     instance
\cite{GasperiniVeneziano,PreBigBang}).  Recently,  forms have appeared
in  attempts   to  model  late-time  acceleration   in  supergravity.  
\cite{supergravity}.  Massive  vectors ($1$-forms) were  considered by
Ford,  who  proposed  a  model  wherein  a  four-dimensional  universe
undergoes inflation driven by a self-interacting vector \cite{Ford}.

The  paper is  organized as  follows.  In  Section  \ref{sec:model} we
present our model.  It is described, both, from the higher-dimensional
point of view, and  from a dimensionally-reduced perspective.  Section
\ref{sec:solutions}  contains details  of the  inflationary solutions,
and  in Section  \ref{sec:perturbations}  we compute  the spectrum  of
primordial  density  fluctuations   generated  during  inflation.   In
Section  \ref{sec:reheating}  we discuss  the  end  of inflation,  and
propose a  way of  stabilizing the radion  whilst avoiding  the severe
constraints on violations of  the equivalence principle.  Two examples
illustrating  phenomenologically realistic  choices of  parameters are
presented  in  Section  \ref{sec:example}.   We  conclude  in  Section
\ref{sec:conclusions}.

\section{The Model}\label{sec:model}
Consider a $(4+d)$-dimensional spacetime, i.e., one which contains $d$
additional  spatial  dimensions. We  will  propose  a mechanism  which
allows  three spatial  dimensions to  inflate, while  keeping  the $d$
remaining ones  small.  Towards this  end, we introduce  an anisotropy
via fields living on the spacetime, rather than through a violation of
$(4+d)$-dimensional diffeomorphism invariance \cite{branes}.  This can
be achieved  by spontaneously giving  an expectation value to  a field
that transforms non-trivially under rotations in $4+d$ dimensions (see
\cite{BrandenbergerVafa} for an alternative possibility).  Such fields
can be  fermionic or bosonic.  Conventional wisdom  suggests that only
bosonic fields may acquire large expectation values.  While this need
not  always be  true \cite{spinors},  we will  adopt such  a viewpoint
here.   This  leaves us  with  one  choice,  vector fields  and  their
generalizations, differential forms.
 
\subsection{Higher-dimensional equations}

Consider    a   (totally    antisymmetric)    $d$-form   $A_{M_1\cdots
  M_d}=A_{[M_1\cdots M_d]}$ minimally coupled to gravity,
\begin{equation}\label{eq:4+daction}
S=\int d^{4+d}x \sqrt{-g}\left[\frac{R}{6}-
  \frac{1}{2(d+1)!}F_{M_1\cdots M_{d+1}}F^{M_1\cdots M_{d+1}}-W(A^2)\right].
\end{equation}
The  $(d+1)$-form $F$  is  the field  strength  of $A$,  $F_{M_1\cdots
  M_{d+1}}=(d+1)\partial_{[M_1}A_{M_2\cdots    M_{d+1}]}$,   and   the
self-interaction  term  $W$ is  an  as  of  yet unspecified  arbitrary
function of $A^2\equiv A_{M_1\cdots M_d}A^{M_1\cdots M_d}$. We work in
units where the four and higher-dimensional Newton's constant is $8\pi
G=3$, and our metric signature is $(-,+,\cdots, +)$.

Although  we have  not  written the  additional  matter terms  in Eq.  
(\ref{eq:4+daction}), we assume that the form does not couple to them.
Hence, the form $A_{M_1\cdots M_d}$  only interacts with gravity. In a
theory of  massless forms, invariance under  the gauge transformations
$\delta  A=dB$,   where  $B$   is  any  $d-1$-form,   guarantees  that
appropriate   components   of    the   form   decouple   from   matter
\cite{Weinberg}.  In our theory, the  form $A$ is not massless, and it
does not couple  to additional matter sources either,  so we shall not
require  gauge invariance.   In fact,  the self-interaction  terms $W$
explicitly violates this symmetry.

Varying the action (\ref{eq:4+daction}) with respect to the metric one
obtains  Einstein's equations  ${G_{MN}=3~T_{MN}}$,  where  the energy
momentum tensor is given by
\begin{equation}\label{eq:EMT}
T_{MN}=\frac{1}{d!}F_{M M_2\cdots M_{d+1}}F_N{}^{M_2\cdots M_{d+1}}+
2~d~W'~A_{M M_2\cdots M_d}A_N{}^{M_2\cdots M_d}-g_{MN}
\left(\frac{F^2}{2(d+1)!}+W\right).
\end{equation}
and a  prime means  a derivative with  respect to $A^2$.   Varying the
action  with respect  to  $A_{M_1\cdots M_d}$  one  obtains the  field
equation
\begin{equation}\label{eq:fieldequation}
\nabla_M  F^{M M_1 \cdots M_d}=2~d!~W'~A^{M_1\cdots M_d},
\end{equation} 
which     implies     the      constraint
$\nabla_{M_1}(\,W'\, A^{M_1 M_2\cdots M_d})=0$.

In  this paper  we are  interested  in cosmological  solutions of  the
equations  of  motion.    Hence,  we  consider  a  $(4+d)$-dimensional
factorizable  spacetime  $g_{MN}$  consisting  of  a  four-dimensional
Friedmann-Robertson-Walker metric $g_{\mu\nu}$ times a $d$-dimensional
compact  internal space  with  metric $G_{mn}$  of constant  curvature
$R^{(d)}$,
\begin{equation}\label{eq:metric}
ds_{4+d}^2\equiv g_{MN}dx^M dx^N=g_{\mu\nu}dx^\mu dx^\nu+b^2 
G_{mn}dx^m dx^n.
\end{equation}
The  coordinates $x^\mu$  label our  four dimensional  world,  and the
coordinates  $x^m$  label  what  we  shall call  the  ``internal''  or
``compact'' space (the dimensions that remain small during inflation).
Once the  four dimensions start  to inflate, the spatial  curvature of
the four-dimensional metric soon becomes negligible.  So, without loss
of generality, we consider a spatially flat universe,
\begin{equation}
ds_4^2\equiv g_{\mu\nu}dx^\mu dx^\nu=-dt^2+a^2(t) \delta_{ij}dx^i dx^j,
\end{equation}
where $t$ is cosmic time and $a(t)$ is the scale factor.  We normalize
the internal space metric by the condition
 \begin{equation}\label{eq:normalization}
\int d^dx \sqrt{G}=1.
\end{equation}
Hence,  the  volume  of  the   internal  space  is  $b^d$.  Then,  the
four-dimensional scalar  $b$ can be  interpreted as the radius  of the
internal space, so we shall call it the radion.

Substituting   the   metric    ansatz   (\ref{eq:metric})   into   the
$(4+d)$-dimensional  Einstein equations and  assuming that  the radion
only depends on time one obtains
\begin{subequations}\label{eq:Einstein}
\begin{eqnarray}
H^2+d H I +\frac{d^2-d}{6}I^2+\frac{R^{(d)}}{6\,b^2}&=&
\rho\\
3H^2+2\dot{H}+\frac{d^2+d}{2}I^2+d\dot{I}+2dHI+\frac{R^{(d)}}{2\,b^2}&=&-3p \\
6H^2+3\dot{H}+\frac{d^2-d}{2}I^2+(d-1)\dot{I}+
3(d-1)HI+\frac{(d-2)R^{(d)}}{2 d\, b^2}&=&-3P.
\end{eqnarray}
\end{subequations}
In  the previous  equations we  have introduced  the  Hubble parameter
$H\equiv  d\log a/dt$  and the  expansion  rate in  the compact  space
$I\equiv d\log b /dt$. A dot means a derivative with respect to cosmic
time $t$.   The energy density $\rho$,  the three-dimensional pressure
$p$,  and   the  pressure  along   the  compact  space  $P$   are  the
corresponding components of the energy momentum tensor
\begin{equation}\label{eq:diagonalEMT}
T^M{}_N=\text{diag}(-\rho,p,p,p,P,\cdots, P).
\end{equation}
Thus, the energy momentum tensor  (\ref{eq:EMT}) has to be diagonal in
order for solutions  of Einsteins equations to exist.   An ansatz that
satisfies this condition is
\begin{equation}\label{eq:ansatz}
A_{M_1\cdots M_d}=\sqrt{G}~\varepsilon_{0123M_1\cdots M_d}~\phi,
\end{equation}
where $\varepsilon_{M_1\cdots  M_{d+4}}$ is totally  antisymmetric and
$\varepsilon_{0\cdots  d+4}=1$.  Consequently,  $A$ has  non-vanishing
components  only  along  the  compact  dimensions.  In  fact,  $A$  is
proportional  to  the  volume  form  in the  compact  space,  and  the
proportionality  factor is the  four-dimensional scalar  $\phi$, which
will turn out to be the  inflaton. In a FRW-universe, the field $\phi$
can only  depend on time.  From  the ansatz (\ref{eq:ansatz})  and Eq. 
(\ref{eq:metric}) the square of $A$ given by
\begin{equation}\label{eq:A^2}
A^2=d!~b^{-2d}~\phi^2,
\end{equation}
i.e. $A^2$ is a combination of the inflaton $\phi$ and the radion $b$.

Substituting Eq.  (\ref{eq:ansatz}) into Eq.  (\ref{eq:EMT}) one finds
that   the   energy   momentum   tensor   is  indeed   of   the   form
(\ref{eq:diagonalEMT}), where  energy density and  pressures are given
by
\begin{subequations}\label{eq:rhoandp}
\begin{eqnarray}
\rho&=&\frac{b^{-2d}}{2}~\dot{\phi}^2+W\\
p&=&\frac{b^{-2d}}{2}~\dot{\phi}^2-W\\
P&=&-\frac{b^{-2d}}{2}~\dot{\phi}^2+2\,W'~A^2-W.
\end{eqnarray}
\end{subequations}
One  can  derive  the  equation   of  motion  for  $\phi$  from  Eqs.  
(\ref{eq:Einstein}) and (\ref{eq:rhoandp})  or directly by considering
Eq. (\ref{eq:fieldequation}) for the internal components,
\begin{equation}\label{eq:motion}
\ddot{\phi}+(3 H-d I)\,\dot{\phi}+2\, d!\, W'\,\phi=0.
\end{equation}
Because of the symmetry of the ansatz, the remaining components of the
equation of motion  are identically satisfied. 

By  definition,   inflation  is  a  stage   of  accelerated  expansion
$\ddot{a}>0$.  From  a four-dimensional perspective,  any inflationary
stage  can explain  the  flatness  and homogeneity  of  the universe.  
However,  in an expanding  universe \cite{GrKhStTu},  only a  stage of
inflation  close to  de  Sitter  yields in  general  the nearly  scale
invariant   spectrum   of   primordial   density   fluctuations   that
observations seem to favor \cite{WMAP}.   Therefore, in order to get a
feeling  of the  constraints  our  $d$-form has  to  satisfy we  shall
consider a de Sitter stage $H=const$.  In addition, we want to explain
why our four-dimensions  are large compared to the  internal ones, and
the simplest way to accomplish that  is to assume that the size of the
internal  dimensions remains  constant. Therefore,  we shall  look for
solutions with static  internal dimensions, $I=\dot{I}=0$.  With these
assumptions Eqs.  (\ref{eq:Einstein}) take the form
\begin{subequations}
\begin{eqnarray}
H^2+\frac{R^{(d)}}{6\,b^2}&=&\rho\label{eq:4+d-Friedmann}  \\
3H^2+\frac{R^{(d)}}{2\,b^2}&=&-3p \\
6H^2+\frac{(d-2)R^{(d)}}{2 d\, b^2}&=&-3P.
\end{eqnarray}
\end{subequations}
Solutions of the previous equations exist if and only if
\begin{subequations}
\begin{eqnarray}
\rho+p&=&0\quad \label{eq:state}\\
2\rho+P&=&\frac{d+2}{6~d}\frac{R^{(d)}}{b^2} \label{eq:weak}.
\end{eqnarray}
\end{subequations}
Eq.  (\ref{eq:state}) is  the familiar inflationary relation satisfied
by  a cosmological  constant or  a  frozen scalar  field.  The  second
condition, Eq.   (\ref{eq:weak}), implies that for  flat or negatively
curved  internal  dimensions  the  null  energy condition  has  to  be
violated \cite{CaGeHoWa,NaSiStTr}.

We restrict  now our  attention to the  energy momentum tensor  of the
$d$-form,  Eqs.    (\ref{eq:rhoandp}).   Then,  Eq.   (\ref{eq:state})
implies  that the  inflaton  is frozen,  $\dot{\phi}=0$.  Although  an
exactly   frozen  field   is  in   general   not  solution   of  Eq.   
(\ref{eq:motion}),  we shall  later see  that a  \emph{nearly} frozen,
slowly-rolling   field  actually  is.    On  the   other  hand,   Eq.  
(\ref{eq:weak}) constraints the form of the interaction $W$,
\begin{equation}\label{eq:condition}
W+ 2 W' A^2=\frac{d+2}{6~d}\frac{R^{(d)}}{b_0^2},
\end{equation}
where  $b_0$  is the  constant  value of  $b$.   The  solution to  the
previous equation is
\begin{equation}\label{eq:potential}
W(A^2)=W_0+W_1\cdot (A^2)^{-1/2},
\end{equation}
where $W_1$  is an arbitrary  integration constant.\footnote{One could
  also  allow   for  functions  that   are  only  \emph{approximately}
  described  by  Eq. (\ref{eq:potential}),  though  for simplicity  we
  shall not explore this possibility  here.}  If $R^{(d)}=0$, then Eq. 
(\ref{eq:condition})   implies   $W_0=0$.    If   $R^{(d)}\neq0$   the
$(4+d)$-dimensional cosmological term $W_0$ is related to the constant
value of $b$ by
\begin{equation}\label{eq:b0}
  b_0=\sqrt{\frac{d+2}{6 d}\frac{R^{(d)}}{W_0}}.
\end{equation}
 
In summary,  if a (nearly)  frozen field $\phi$  is a solution  of the
equations  of  motion, there  exist  solutions  where four  dimensions
inflate and the  where radius of the internal  dimensions is constant. 
If the compact internal dimensions are flat, $R^{(d)}=0$, any constant
value of $b$  is possible.  If the internal  dimensions are positively
curved,  $R^{(d)}>0$,   a  positive  $(4+d)$-dimensional  cosmological
constant  $W_0$ is required,  whereas if  the internal  dimensions are
negatively curved, $R^{(d)}<0$, a negative cosmological constant $W_0$
is needed to stabilize the radion during inflation.

Compact spaces  of constant negative  curvature can be  constructed by
acting  on the  $d$-dimensional  hyperbolic plane  $H^d$  with a  free
discrete isometry  group of the space  \cite{Thurston}.  For $d\geq3$,
these compact  hyperbolic manifolds are  ``rigid'', in the  sense that
the radius is the only massless  moduli they admit.  In that case, our
metric ansatz for the metric in the internal space is the most general
one.  The application of compact hyperbolic manifolds to cosmology has
been pioneered by Starkman  and collaborators (see \cite{KaMaStTr} and
references  therein). Recently,  compact hyperbolic  spaces  have also
received  attention  in  the  context  of  cosmological  solutions  in
supergravity \cite{supergravity}.

Below we shall see that a  nearly frozen field is indeed a solution of
the  equations  of  motion,  completely analogous  to  a  conventional
slow-roll inflationary regime.  We shall also show that if the compact
space  is  positively  curved,  solutions  with  constant  radion  are
unstable.  In order to understand these properties though, it is going
to be convenient to work with the dimensionally reduced action.

\subsection{Dimensionally reduced action}
In the cosmological setting we  have been dealing with, the radion $b$
and the  inflaton $\phi$ only depend  on time. In  particular, $b$ and
$\phi$  do not  depend  on  the internal  coordinates  $x^m$.  If  the
different fields  in the action (\ref{eq:4+daction}) do  not depend on
the  internal space,  it is  possible to  integrate over  the internal
space and obtain a four dimensional action that describes the dynamics
of  the four-dimensional  scalars $b$  and $\phi$.   Substituting  Eq. 
(\ref{eq:ansatz})  into the  Lagrangian of  Eq.  (\ref{eq:4+daction}),
and integrating over the internal space we obtain
\begin{equation}\label{eq:Jordan-frame}
  S=\int d^4x \sqrt{-g} \,b^d\left[\frac{R^{(4)}}{6}
    +\frac{R^{(d)}}{6\,b^2}
    +\frac{d^2-d}{6}\frac{\partial_\mu b \,\partial^\mu b}{b^2}
    -\frac{1}{2}\frac{\partial_\mu \phi \,\partial^\mu \phi}{b^{2d}}
    -W(A^2)\right],
\end{equation}  
where we  have used Eq. (\ref{eq:normalization}) and  $A^2$ stands for
the r.h.s  of Eq.  (\ref{eq:A^2}).  $R^{(4)}$  is the four-dimensional
scalar  curvature  and wherever  the  metric  tensor  is involved  the
four-dimensional  metric $g_{\mu\nu}$ is  to be  used.  Note  that the
kinetic terms of the different  fields are not in canonical form.  For
convenience, we  would like to  work with canonical kinetic  terms for
the metric and the radion, so we shall rename those fields,
\begin{equation}
  g_{\mu\nu}\equiv b^{-d}\widetilde{g}_{\mu\nu}
  \quad\text{and} \quad
  b\equiv e^{\sigma}.
\end{equation}
Plugging the  last expressions into Eq.  (\ref{eq:Jordan-frame}) we
get
\begin{equation}\label{eq:Einstein-frame}
  S=\int d^4x \sqrt{-g} \left[\frac{R^{(4)}}{6}
    -\frac{d^2+2d}{12}\partial_\mu \sigma \,\partial^\mu \sigma
    -\frac{1}{2}\frac{\partial_\mu \phi \,\partial^\mu \phi}{e^{2d\sigma}}
    +e^{-(d+2)\sigma}\frac{R^{(d)}}{6}-e^{-d\sigma}W(A^2)\right],
\end{equation}
where all quantities  refer to the tilded metric  (we have dropped the
tilde).   Because  for the  solutions  we  are  interested in  $b$  is
(nearly)  constant, the  behavior of  the scale  factor in  the Jordan
frame     (\ref{eq:Jordan-frame})    and     the     Einstein    frame
(\ref{eq:Einstein-frame})  is  essentially  the  same. Let  us  stress
though,  that in  any  case all  ``conformal''  frames are  physically
equivalent, since  physical predictions should  not depend on  the way
fields are named (see for instance \cite{varying}).

\begin{figure}
\includegraphics{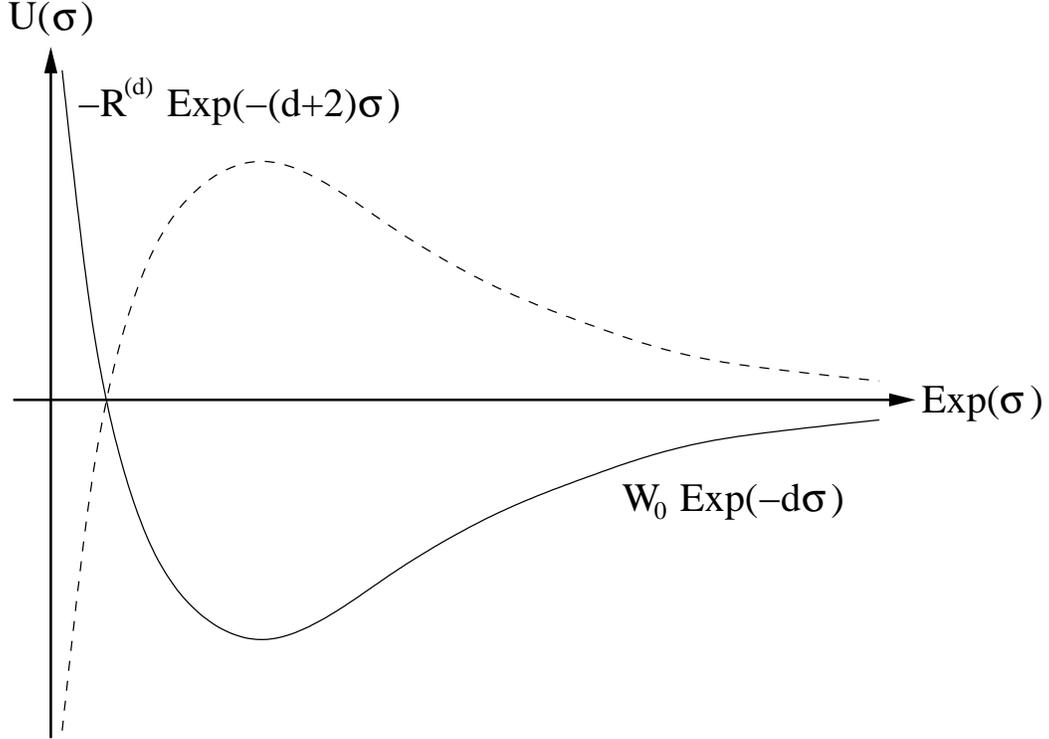}
\caption{A plot of the radion potential, Eq. (\ref{eq:radionpotential}),
  for  different signs of  the compact  space curvature  $R^{(d)}$. In
  order     for    the    potential     to    have     an    extremum,
  $\text{sgn}(R^{(d)})=\text{sgn}(W_0)$.   The   extremum  is  then  a
  minimum   if  $W_0<0$.   For   $R^{(d)}=0$,  the   radion  potential
  identically vanishes.\label{fig:radionpotential}}
\end{figure}

The action (\ref{eq:Einstein-frame})  describes two interacting scalar
fields  $\phi$ and  $\sigma$.   Substituting Eq.   (\ref{eq:potential})
into Eq.   (\ref{eq:Einstein-frame}) and using  Eq. (\ref{eq:A^2}) one
obtains a potential term
\begin{equation}
  U(\sigma)+V(\phi)\equiv-e^{-(d+2)\sigma}\frac{R^{(d)}}{6}+
  e^{-d\sigma}W_0+\frac{W_1}{\sqrt{d!}} \frac{1}{\phi},
\end{equation}
where  by  ``potential''  we  mean  the  sum of  those  terms  in  the
Lagrangian  that do  not contain  derivatives of  the fields,  with an
overall minus  sign.  Remarkably, this potential does  not involve any
coupling between the radion $\sigma$ and the inflaton $\phi$, which is
part of the origin  of the condition (\ref{eq:condition}).  Therefore,
as far  as the potential is  concerned, the two  fields are decoupled,
and  we can  write down  the potential  terms for  the radion  and the
inflaton separately.  For  non-vanishing internal curvature the radion
potential is
\begin{equation}\label{eq:radionpotential}
U(\sigma)=-e^{-(d+2)\sigma}\frac{R^{(d)}}{6}+e^{-d\sigma}W_0,
\end{equation}
which has  an extremum at $\sigma_0=\log  b_0$, and $b_0$  is given by
Eq.    (\ref{eq:b0}).   The  second   derivative  of   $U(\sigma)$  at
$\sigma_0$ is
\begin{equation}
  m_\sigma^2\equiv\frac{6}{d^2+2d}\frac{d^2 U}{d\sigma^2}\Big|_{\sigma_0}=
  -\frac{2e^{-(d+2)\sigma_0}}{d}\, R^{(d)}.
\end{equation}
Therefore,  for $R^{(d)}>0$  (i.e. $W_0>0$)  a solution  with constant
$b=b_0$ is unstable, whereas for $R^{(d)}<0$ (i.e. $W_0<0$) a solution
with constant  $b=b_0$ is  perfectly stable. In  the latter  case, the
minimum   of  $V(\sigma)$   occurs  at   a  negative   value   of  the
four-dimensional cosmological constant
\begin{equation}\label{eq:lambda}
  U(\sigma_0)=
  \frac{2}{d+2}\left(\frac{d+2}{6 d}\frac{R^{(d)}}{W_0}\right)^{-d/2}
W_0.
\end{equation} 
If the internal curvature vanishes  (i.e. $W_0=0$), so does the radion
potential.  A constant $b$ is then marginally stable, and any value of
$b_0\equiv  e^{\sigma_0}=const$  is a  solution  of  the equations  of
motion. A plot  of the potential for the  different signs of $R^{(d)}$
is shown in Fig. \ref{fig:radionpotential}.
    
\section{Inflationary solutions}\label{sec:solutions}

Our  next goal  is to  study  the inflationary  solutions with  nearly
constant radion  that we have hinted  in the previous  sections and to
investigate some of  their properties. Let us begin  with the inflaton
equation of motion,
\begin{equation}\label{eq:phimotion}
  \ddot{\phi}+(3H-2d\,\dot{\sigma})\,\dot{\phi}
  +e^{2d\sigma}\frac{dV}{d\phi}=0,
\end{equation} 
where the inflaton potential is given by
\begin{equation}\label{eq:Vofphi}
  V(\phi)=\frac{W_1}{\sqrt{d!}}\frac{1}{\phi}.
\end{equation}
Inverse power-law potentials such as the one in Eq.  (\ref{eq:Vofphi})
are known  to admit  ``slow-roll'' inflationary solutions,  since they
satisfy  the  slow-roll  conditions  for  large values  of  the  field
\cite{MuFeBr}.   We will  show that  during slow-roll,  the  radion is
nearly  constant and  the field  acceleration is  negligible.  At this
point let us just assume
\begin{equation}\label{eq:sr1}
  \frac{\dot{\sigma}}{H}\ll 1 \quad \text{and} \quad
    \ddot{\phi}\ll 3H\dot{\phi},
\end{equation}
which implies  that the speed  of the inflaton  field is given  by the
slow-roll expression
\begin{equation}\label{eq:srphidot}
  \dot{\phi}=-\frac{ e^{2d\sigma}}{3H}\frac{dV}{d\phi}.
\end{equation}
Once we  have found the  solutions of the  equations of motion  in the
slow-roll  regime, we  will show  that the  assumptions made  in their
derivation hold.  Note that  the first condition in Eq. (\ref{eq:sr1})
means    that     the    radion    is     nearly    constant    during
slow-roll.\footnote{Quantum  fluctuations are  also responsible  for a
  growth          in           the          expectation          value
  $\langle\sigma^2-\langle\sigma\rangle^2\rangle\approx   H^3\, t$
  during inflation \cite{Vilenkin}.  We  shall assume that this effect
  is negligible.}

The radion equation of motion is 
\begin{equation}
  \ddot{\sigma}+3H\dot{\sigma}+\frac{6}{d+2} e^{-2d\sigma}\dot{\phi}^2
  +\frac{6}{d^2+2d}\frac{dU}{d\sigma}=0,
\end{equation}
where  $U(\sigma)$  is given  by  Eq. (\ref{eq:radionpotential}).   If
$R^{(d)}$ is negative, $U(\sigma)$ has a minimum at $\sigma_0$. Let us
assume  that  the radion  remains  in  the  vicinity of  the  minimum,
$\sigma\approx\sigma_0$,   which  implies   $dU/d\sigma\approx0$.   If
$R^{(d)}$ is zero the  potential $U(\sigma)$ identically vanishes, and
so does $dU/d\sigma$.  Then, if the slow roll assumption
\begin{equation}\label{eq:sr2}
  \ddot{\sigma}\ll 3H\dot\sigma
\end{equation}
is satisfied, the speed of the radion turns to be
\begin{equation}\label{eq:srsigmadot}
  \dot{\sigma}=-\frac{2\,e^{-2d\sigma}}{(d+2)H}\,\dot{\phi}^2.
\end{equation}
Note  that  $\dot{\sigma}$ is  negative,  i.e.   the extra  dimensions
contract.  Because $\dot{\sigma}\equiv \dot{b}/b$, the first condition
in (\ref{eq:sr1})  means that the extra dimensions  evolve much slower
than the inflating ones, as desired.
 
From       the       higher-dimensional       Friedmann       equation
(\ref{eq:4+d-Friedmann})      or      directly      from      Eq.      
(\ref{eq:Einstein-frame}),  the  four-dimensional  Friedmann  equation
reads
\begin{equation}\label{eq:Friedmann}
  H^2=\frac{e^{-2d\sigma}}{2}\dot{\phi}^2
  +\frac{d^2+2d}{12}\dot{\sigma}^2+V(\phi)+U(\sigma),
\end{equation}
where  $V(\phi)$ and  $U(\sigma)$  are respectively  the inflaton  and
radion potentials.  If  $R^{(d)}=0$, $U(\sigma)$ identically vanishes. 
Inflation occurs if the energy density in the universe is dominated by
potential energy of the scalars. Assuming that
\begin{equation}\label{eq:sr3}
  \frac{d^2+2d}{12}\dot{\sigma}^2\ll
  \frac{e^{-2d\sigma}}{2}\dot{\phi}^2\ll V(\phi)+U(\sigma)
\end{equation}
the Friedmann equation  reads
\begin{equation}\label{eq:srFriedmann}
  H^2\approx V(\phi)+U(\sigma).
\end{equation}
Inserting  (\ref{eq:srFriedmann}) into  Eqs.   (\ref{eq:srphidot}) and
(\ref{eq:srsigmadot})   one   can   express  $H$,   $\dot{\phi}$   and
$\dot{\sigma}$  entirely in  terms of  the  values of  the radion  and
inflaton  fields. The  conditions  (\ref{eq:sr1}), (\ref{eq:sr2})  and
(\ref{eq:sr3})  then  reduce, up  to  factors  of  order one,  to  the
slow-roll conditions
\begin{equation}\label{eq:srparameters}
  \epsilon\equiv e^{2d\sigma}\left(\frac{dV/d\phi}{V(\phi)
      +U(\sigma)}\right)^2\ll  1,
  \quad
  \eta\equiv e^{2d\sigma}\frac{d^2 V/d\phi^2}{V(\phi)+U(\sigma)}\ll 1,
\end{equation}
which up to the exponential of $\sigma$ are the conventional slow-roll
conditions of  single-field slow-roll inflation  \cite{Linde}. For the
inflaton potential (\ref{eq:Vofphi}), both conditions are satisfied in
the inflaton lies in the range
\begin{equation}\label{eq:range}
\frac{\exp(d\sigma)}{\sqrt{18}}
\lesssim\phi\lesssim-\frac{1}{\sqrt{d!}}\frac{W_1}{U(\sigma)}. 
\end{equation}
Note that for flat internal dimensions ($U(\sigma)=0$) the upper limit
is  shifted to  infinity.   For  simplicity, we  shall  assume in  the
following that  during inflation, even  if $R^{(d)}\neq 0$,  the field
$\phi$ is much smaller than  the upper limit in Eq.  (\ref{eq:range}). 
Then, $V(\phi)\gg U(\sigma)$ and $U(\sigma)$ can be neglected.

The  slow-roll solutions  we  have discussed  are  attractors for  the
evolutions of the field \cite{Belinsky}.  This means that even if they
are initially  not satisfied, the  equations of motion will  drive the
field to  a regime  where they are  \cite{chaotic}. Therefore,  in our
scenario cosmic evolution is  quite insensitive to initial conditions. 
During slow-roll, the relative change of the Hubble parameter
during a Hubble time is small,
\begin{equation}
  \frac{\dot{H}}{H^2}=-\frac{\epsilon}{6},
\end{equation}
i.e. during slow-roll the universe inflates almost like in a de Sitter
stage. At  the same  time, the radion  and inflaton fields  are nearly
frozen,
\begin{equation}
  \frac{\dot{\sigma}}{H}=-\frac{2\,\epsilon}{9(d+2)},
  \quad e^{-d\sigma}\frac{\dot{\phi}}{H}=-\frac{\sqrt{\epsilon}}{3}
\end{equation}
although  the inflaton  evolves much  faster than  the  radion.  Here,
$\epsilon$   is    the   slow-roll   parameter   defined    in   Eq.   
(\ref{eq:srparameters}).   Thus,  to leading  order  in the  slow-roll
approximation, we  can assume that  the inflaton slowly  changes while
the radion  remains frozen at $\sigma=\sigma_0$.   Here, $\sigma_0$ is
the  minimum of  the radion  potential if  the compact  dimensions are
negatively curved,  or the  initial value of  the radion if  the extra
dimensions are flat.

In  order  for inflation  to  successfully  explain  the flatness  and
homogeneity and the visible universe, it is necessary for inflation to
last  more than around  60 e-folds  (the exact  number depends  on the
unknown  details of  reheating \cite{HuiDodelson}).   We  will discuss
ways to  terminate inflation end in  Section \ref{sec:reheating}.  For
our  present purposes  it  will  suffice to  assume  that because  the
inflaton    potential    deviates    from    its    functional    form
(\ref{eq:Vofphi}), inflation ends when  the inflaton reaches the value
$\phi_0$,   i.e.   when  the   form  reaches   the  value   $A_0^2=d!  
\exp(-2d\sigma_0)\phi_0^2$.  Let us then assume that at some moment of
time during inflation the value the squared form is $A^2$.  The number
of e-folds of inflation between that time and the end of inflation is
\begin{equation}\label{eq:efolds}
  N\equiv\log\frac{a_0}{a}\approx
  \frac{3}{2\,d!}(A_0^2-A^2),
\end{equation}
where we  have assumed slow-roll  in the derivation.   Given $N\approx
60$  and   because  the  inflationary   regime  is  limited  by   Eq.  
(\ref{eq:range}), Eq. (\ref{eq:efolds}) constrains the possible values
of $A_0^2$.

\section{Perturbations}\label{sec:perturbations}
The most important consequence of a stage of quasi de Sitter inflation
is the generation  of a nearly scale invariant  spectrum of primordial
density perturbations  \cite{Mukhanov}.  In the  simplest inflationary
models,  inflation  is  driven  by  a single  scalar  field,  and  its
fluctuations are responsible for  the adiabatic primordial spectrum of
fluctuation that observations seem to favor.  The presence of a second
(light) scalar field during  inflation is potentially dangerous, since
it   could   lead  to   entropy   perturbations   (see  for   instance
\cite{MukhanovSteinhardt}).

In  our scenario,  inflation is  driven by  the single  inflaton field
$\phi$, but there is an additional field, the radion $\sigma$.  If the
compact  dimensions  are flat,  to  leading  order  in slow-roll,  the
potential  for $\sigma$  vanishes.  Because  of that,  fluctuations of
$\sigma$ do not  contribute to fluctuations in the  energy density, so
that   no   entropic   component   is   generated   during   slow-roll
\cite{MukhanovSteinhardt}.  If  the compact dimensions  are negatively
curved, the field $\sigma$ is massive.  If its mass is bigger than the
Hubble factor  during inflation, quantum fluctuations  of $\sigma$ are
suppressed,  so no  entropy component  is generated  either.   In both
cases, we  can therefore regard $\sigma$  as a constant  and study the
perturbations solely due to the inflaton $\phi$.

The power  spectrum $\mathcal{P}(k)$ is  a measure of the  mean square
fluctuations  of  the  generalized  Newtonian  potential  in  comoving
distances  $1/k$.   During  radiation   domination,  it  is  given  by
\cite{MuFeBr}
\begin{equation}\label{eq:power}
  \mathcal{P}(k)=\frac{e^{-2d\sigma_0}}{\pi^2}
  \frac{V^3}{V^2_{,\phi}}\Bigg|_{k=aH},
\end{equation}
where $V$  is the potential (\ref{eq:Vofphi}) and  $k=aH$ denotes that
the r.h.s.  of  the equation has to be evaluated at  the time the mode
crosses  the Hubble  radius.   The factor  $e^{-2d\sigma_0}$ shows  up
because we  work with  a non-canonically normalized  inflaton. Because
the radion  evolves much  slower than the  radion, we assume  that the
radion is constant.

The  amplitude of  the power  spectrum is  approximately equal  to the
squared  amplitude  of  the  temperature fluctuations  in  the  cosmic
microwave   background,  $\delta   T/T\approx   10^{-5}$.   Evaluating
$\mathcal{P}(k)$ for a mode that crosses the Hubble radius $N$ e-folds
before the  end of inflation, Eq.  (\ref{eq:efolds}),  we hence obtain
the constraint
\begin{equation}\label{eq:temperature}
  \mathcal{P}=\frac{e^{-d\sigma_0}}{\pi^2}\frac{W_1}{\sqrt{d!}}
  \,\sqrt{\frac{A_0^2}{d!}-\frac{2 N}{3}}\approx 10^{-10}.
\end{equation}  
Observe the exponential  suppression with increasing $d\sigma_0$.  The
(nearly constant) slope  of the power spectrum is  parametrized by the
spectral index $n_s$, where
\begin{equation}
  n_s-1\equiv \frac{d\log \mathcal{P}}{d\log k}.
\end{equation} 
Current     observations     imply     the    limits     $0.9<n_s<1.1$
\cite{BaggerLeeMarfatia}.  Substituting Eq.  (\ref{eq:power}) into the
previous definition  and evaluating it  $N$ e-folds before the  end of
inflation, Eq. (\ref{eq:efolds}), we obtain
\begin{equation}\label{eq:spectral}
  n_s-1=\frac{1}{3 A_0^2/d!-2N}<10^{-1}.
\end{equation}
Note  that  the  quantity  in  the denominator  is  always  positive.  
Therefore,  the  power  spectrum  is blue  ($n_s>1$).   Cosmologically
relevant scales typically left the horizon about $N\approx 60$ e-folds
before  the end  of  inflation,  though this  number  varies with  the
details of  reheating \cite{HuiDodelson}. We show below  that in order
to satisfy constraints  on the universality of free  fall, $A_0^2$ has
to  be large,  $A_0^2>d!\cdot 10^7$.   Hence, in  general  one expects
$n_s\approx 1$.  In Section  \ref{sec:example} we deal with a concrete
example where the different  parameters ($W_1$, $\sigma_0$, $d$, etc.) 
in phenomenologically viable models are specified.

\section{Reheating and Radion Stabilization}\label{sec:reheating}

\begin{figure}
  \includegraphics{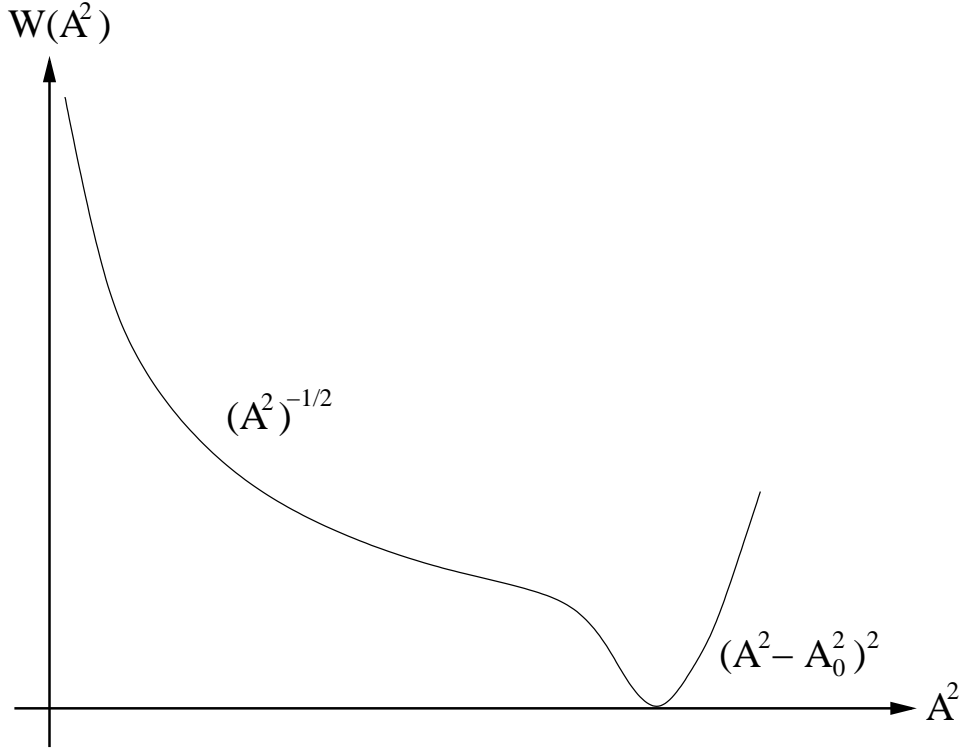}
  \caption{A plot of the form self-interaction $W$.
    \label{fig:inflatonpotential}}
\end{figure}

In the conventional four-dimensional models of inflation, the universe
is reheated  when inflation ends  and the inflaton  starts oscillating
around the minimum of its  potential \cite{KoLiSt}.  In our model, the
inflaton   potential   is  given   by   Eq.   (\ref{eq:Vofphi}).    If
$R^{(d)}<0$,   inflation   ends   when   $\phi$  reaches   the   value
$W_1/(d!\,U(\sigma_0))$.  At  that value  of the field,  the potential
energy vanishes, and beyond that value the potential becomes negative.
The  cosmological evolution  of  a scalar  field  with such  effective
potentials  was studied  in  \cite{FeFrKoLi}.  It  was  found that  in
general, once  the potential becomes  negative, the universe  enters a
phase of contraction  that ends in a singularity.   This would prevent
the  universe from  reheating, thus  invalidating our  model.   If the
internal space  is flat, $R^{(d)}=0$, once the  slow-roll condition is
satisfied, it is never violated. In that case inflation never ends. In
both  cases we  have to  assume that  around some  value of  the field
$\phi_0$   the   potential   deviates   from   its   form   in   Eq.   
(\ref{eq:Vofphi})   and  is  described   by  a   different  functional
form\footnote{For an alternative mechanism to end inflation in a model
  with  $V(\phi)\propto 1/\phi$, see  \cite{BrandenbergerZhinitsky}.}. 
This  is the  case  if  at some  point  the self-interaction  $W(A^2)$
develops    a   minimum    at   $A_0^2$,    as   shown    in    Fig.   
\ref{fig:inflatonpotential}.  Around the minimum, the function $W$ can
be expanded as
\begin{equation}\label{eq:minimumpot}
  W(A^2)\approx V_0+\frac{\lambda}{8} (A^2-A_0^2)^2,
\end{equation}
where  $V_0$  is a  cosmological  term  and  $\lambda$ is  a  coupling
parameter.   From  the  dimensionally  reduced  point  of  view,  Eq.  
(\ref{eq:Einstein-frame}), this yields a radion and inflaton potential
\begin{equation}\label{eq:flatpotential}
  V(\sigma,\phi)=-e^{-(d+2)\sigma}\frac{R^{(d)}}{6}
  +e^{-d\sigma}\left[V_0+\frac{\lambda}{2}
    \left(d!\, e^{-2d\sigma}\phi^2-A_0^2\right)^2\right].
\end{equation}
Again,  if  $R^{(d)}<0$,  the   potential  has  a  stable  minimum  at
$b_0\equiv  e^{\sigma_0}$  given  by  Eq.  (\ref{eq:b0}),  with  $W_0$
replaced by $V_0$. Even if  $W$ has a minimum, a negative cosmological
term is needed to stabilize  the radion if the internal dimensions are
negatively  curved   \cite{NaSiStTr}.   Although  the   value  of  the
cosmological  term during  inflation and  after the  end  of inflation
might have changed, it can do  so only if the value of $\sigma_0$ that
minimizes the  effective potential changes  significantly. Because one
of our main  goals was to study inflationary  solutions where $\sigma$
is constant, we shall not  consider this case anymore and proceed with
flat internal dimensions.

If $R^{(d)}=0$, we  shall assume $V_0=0$, which amounts  to tuning the
higher-dimensional cosmological  constant to zero.   During inflation,
$\phi$ evolves while $\sigma$  stays constant.  When $A^2$ reaches the
vicinity  of  $A_0^2$ inflation  ends,  and  the  fields approach  the
minimum at $A_0^2$.   Because we assume that the  end of inflation and
the minimum of  $W$ are not far apart we can  assume that $\sigma$ and
$\phi$ do not change significantly. Then
\begin{equation}\label{eq:A0}
  A_0^2 \approx d!~e^{-2d\sigma_0} \phi_0^2,
\end{equation}
where $\sigma_0$  is the constant  value of $\sigma$  during inflation
and $\phi_0$ is the value of $\phi$ at the end of inflation.

Although there are reheating  models where the inflaton potential does
not oscillate around a minimum \cite{FeKoLi}, we would like one of our
fields  to oscillate  around the  minimum of  its potential,  i.e.  we
would   like   one  of   them   to   get   a  mass.    The   potential
(\ref{eq:minimumpot}) has  a minimum at  $A^2=A_0$.  Let us  denote by
$\sigma_0$  and $\phi_0$  the values  of $\sigma$  and $\phi$  at that
minimum, and let us consider fluctuations around those values,
\begin{eqnarray}
  \phi=\phi_0+e^{d\sigma_0}\,\delta\phi\quad   \text{and}\quad  
  \sigma=\sigma_0+\sqrt{\frac{6}{d^2+2d}}~\delta\sigma.
\end{eqnarray}
Substituting  the previous  definitions  into the  Lagrangian  of Eq.  
(\ref{eq:Einstein-frame})  and  expanding   to  second  order  in  the
fluctuations one  gets a coupled system of  two canonically normalized
fields  $\delta\sigma$ and  $\delta\phi$.  Let  us introduce  the also
canonically normalized  fields $\chi_0$  and $\chi_1$, defined  by the
relations
\begin{equation}\label{eq:fields}
 \delta\phi=\cos \theta \,\chi_0+\sin \theta\, \chi_1,\quad
  \delta\sigma=-\sin \theta \, \chi_0+\cos\theta \, \chi_1,
\end{equation}
where
\begin{equation}\label{eq:theta}
\sin^2\theta=\left(1+\frac{6d}{d+2}\frac{A_0^2}{d!}\right)^{-1}.
\end{equation}
The fields  $\chi_0$ and $\chi_1$  diagonalize the mass matrix  of the
fluctuations.  Because the  potential Eq.  (\ref{eq:minimumpot}) has a
flat direction, the  scalar $\chi_0$ turns to be  massless.  The field
$\chi_1$ is massive, and its mass is given by
\begin{equation}
  m^2_{\chi_1}=\frac{\lambda\, d!\, 
    e^{-d\sigma_0}A_0^2}{\sin^2\theta}.
\end{equation}
We shall show below that in order to satisfy experimental restrictions
related to the apparent universality of free fall, we have to consider
the   small  mixing  angle   limit  $\theta\ll   1$.   In   that  case
$\delta\sigma\approx \chi_1$  is massive and $\delta\phi\approx\chi_0$
is massless.   Therefore, at the  end of inflation, the  radion starts
oscillating around  the minimum of  its potential, while  the inflaton
remains essentially constant.   Note that for reheating to  work it is
not crucial  that the field that  drives reheating is the  same as the
one  that   drives  inflation.   The   inhomogeneities  seeded  during
inflation  can be  transferred to  the  decay products  of the  radion
because  the terms  that couple  the oscillating  radion to  matter in
general contain metric  fluctuations (see \cite{reheating} for related
ideas). There is  nevertheless a potential challenge our  model has to
face.  The radion is massless during inflation, but becomes massive at
the  end  of  inflation  and   oscillates  around  its  minimum  at  a
non-vanishing    $\sigma_0$.     Hence,     as    pointed    out    in
\cite{FinelliBrandenberger}  it is  possible that  metric fluctuations
are parametrically amplified during  the reheating process.  This will
happen  if  long-wavelength modes  lie  in  a  resonance band  of  the
equation  that describes the  evolution of  the radion  perturbations. 
Because in our case the  mass (and therefore the oscillation frequency)
of  the  radion can  be  freely  adjusted  by changing  the  parameter
$\lambda$, one might avoid such resonances.

Once  inflation  has ended  and  the  universe  has been  heated,  the
universe  evolves according  to the  standard Big-Bang  scenario.  The
presence  of  the  massless  scalar field  $\chi_0$  that  generically
couples to matter  could yield however to violations  of several tests
on  the couplings  of matter  to gravity  and on  the validity  of the
equivalence   principle  \cite{Damour,Sean}.    Presently,   the  most
stringent restrictions  arise from experiments on  the universality of
free fall.   Upon the dimensional reduction  of the higher-dimensional
action (\ref{eq:4+daction})  (including the  matter terms we  have not
written   down),   one  expects   the   radion   to   couple  to   the
four-dimensional  fields  with different  powers  of $e^\sigma$.   For
instance, a  Maxwell term $F_{MN}F^{MN}$  in Eq.  (\ref{eq:4+daction})
leads   to  the   term  $e^{d\sigma}F_{\mu\nu}F^{\mu\nu}$   in   Eq.  
(\ref{eq:Einstein-frame})  [by  $F$ we  now  mean the  electromagnetic
field strength, not the field  strength of our form $A$].  If $\chi_0$
is a massless canonically normalized scalar, the coupling strength
\begin{equation}\label{eq:alpha}
  \alpha\equiv\frac{\partial \log e^{-d\sigma}}{\partial \chi_0}=
  -\sqrt{\frac{6d}{d+2}}\frac{\partial \delta\sigma}{\partial\chi_0}
\end{equation} 
is  severely restricted  by experiments  on the  universality  of free
fall, where the undetected  differential acceleration of two bodies of
different   composition   puts   the  limit   $\alpha^2\leq   10^{-7}$
\cite{Damour}. Using Eqs.  (\ref{eq:fields}), (\ref{eq:theta}) and the
last limit we find
\begin{equation}\label{eq:freefall}
  \frac{A_0^2}{d!}\geq 10^7.
\end{equation}
It then follows that the  field $\delta\sigma$ points in the direction
of the  massive field $\chi_1$, whereas the  field $\delta\phi$ points
along the  massless direction $\chi_0$,  Eq.  (\ref{eq:fields}).  Note
that since in  our model the form $A_{M_1\cdots  M_d}$ only couples to
gravity,  in the dimensionally  reduced action  the field  $\phi$ only
interacts  gravitationally.    Therefore  there  are   no  constraints
originating from the massless field $\chi_0\approx \delta\phi$.

\section{Two Examples}\label{sec:example}
In the  previous Sections we  have formulated the conditions  that our
scenario  has  to  satisfy  in  order  to  provide  for  a  successful
inflationary scenario,  but we have  not verified whether they  can be
satisfied  at all.   In  the following,  we  shall fill  this gap  and
comment on  the nature of  those conditions by providing  two explicit
examples.  Essentially, the conditions our  model has to obey have two
different  origins: Constraints related  to inflation  and constraints
related to radion  stabilization after the end of  inflaton.  Our main
goal consisted of formulating an inflationary scenario where a certain
number  of  dimensions  remain  small  while  four  dimensions  become
exponentially  large.  As  a bonus,  we have  also proposed  a  way to
stabilize the radion after the end of inflation. Because inflation and
radion stabilization  are in principle two different  issues, we shall
deal with the two sources of constraints separately.

Let  us  first  discuss  the  constraints  associated  to  inflation.  
Inflation  has   to  last  longer   than  about  $60$  e-folds,   Eq.  
(\ref{eq:efolds}), it has to account  for the correct amplitude of the
temperature anisotropies  Eq. (\ref{eq:temperature}) and  must explain
the   nearly   scale  invariance   of   the   power   spectrum,  Eq.   
(\ref{eq:spectral}).  In  addition, after the  end of inflation  it is
desirable that the  universe is reheated when one  of our scalars, the
radion or  the inflaton starts  oscillating around the minimum  of its
potential.   Let  us  assume  that  the compact  dimensions  are  flat
($R^{(d)}=0$). String theories require the existence of $6$ additional
spatial   dimensions, hence, we shall set
\begin{equation}\label{eq:dimensions}
  d=6.
\end{equation}
Because the  compact dimensions are flat,  we have to  pick $W_0=0$ in
order to satisfy Eq. (\ref{eq:condition}).  This choice corresponds to
tuning  the cosmological  constant to  zero.  The  parameter  $W_1$ is
constrained by Eq. (\ref{eq:temperature}).  Notice that the larger the
extra  dimensions, the larger  $W_1$ can  be.  Thus,  extra dimensions
might ``explain''  the smallness of the  temperature anisotropies.  As
mentioned in  the introduction, one expects all  spatial dimensions to
be  Planck  sized initially.   In  fact, for  a  smaller  size of  the
universe, our classic description is likely to break down.  Therefore,
we can  safely rely  on classical general  relativity if  for instance
$\sigma_0\approx \log 10$, i.e.
\begin{equation}\label{eq:size}
  b_0\equiv e^{\sigma_0} \approx 10.
\end{equation}
With   this   choice   of   $\sigma_0$,  the   slow   roll   condition
(\ref{eq:range})  is  satisfied  for   $A^2\gtrsim  40$.   Then,  Eq.  
(\ref{eq:efolds})  guarantees than  the maximal  number of  e-folds is
more than enough if we choose
\begin{equation}
  A_0\approx 220.
\end{equation}
which  corresponds to  $N_{max}\approx 100$.  Substituting  the former
values  into Eq.   (\ref{eq:temperature}) for  a scale  that  left the
horizon $N=60$ e-folds before the end of inflation we obtain
\begin{equation}
W_1\approx 10^{-2},
\end{equation}
a surprisingly ``natural'' number. Let  us point again that its origin
is the exponential suppression  of the temperature anisotropies due to
the internal dimensions.  With  these numbers the spectral index turns
to be
\begin{equation}
  n_s-1\approx 10^{-2},
\end{equation}
quite close  to scale  invariance and  well within the  limits of  Eq. 
(\ref{eq:spectral}).

If  one   considers  in  addition   the  constraint  imposed   by  the
universality of free fall, the numbers are less appealing, though they
are in no way more ``unnatural'' than in other models. Let us keep the
same  number  and   size  of  the  internal  flat   dimensions,  Eqs.  
(\ref{eq:dimensions})  and (\ref{eq:size}).   Then, the  limit in  Eq. 
(\ref{eq:freefall})  is  satisfied  if   we  set  $A_0\approx  10^5$.  
Following  the  same  steps  as  before  translates  into  $W_1\approx
10^{-5}$  and $n_s=1  $.  Therefore,  we see  that in  our  model, the
source of small parameters is not the smallness of the CMB temperature
fluctuations, but the severe limits on the universality of free fall.

\section{Summary and Conclusions}\label{sec:conclusions}
We  have  shown that  Einstein  gravity  in  $(4+d)$ dimensions,  when
coupled to a self-interacting  $d$-form, possesses solutions which are
inflationary  attractors.  These  spacetimes  exhibit an  inflationary
phase in $4$  dimensions, whilst being almost static  in the $d$ extra
dimensions.   Such solutions exist  both for  flat and  for negatively
curved  compact extra  dimensions. However,  in our  minimal scenario,
solutions  with  positively curved  extra  dimensions  and a  constant
radion are unstable.

An attractive  feature of our  model is that  from the viewpoint  of a
four-dimensional observer, the inflaton and the radion are very weakly
coupled.  Therefore, a separate mechanism  is not required in order to
stabilize  the  radion. Instead,  a  bulk  cosmological constant  will
suffice.  More generally, this feature of our model guarantees that if
the radion is  stabilized by a separate mechanism,  inflation will not
destabilize it.

Furthermore,  we are  able to  account  for the  near scale  invariant
spectrum,  and amplitude  of perturbations  with  surprisingly natural
input parameters.   The compact extra dimensions  cause an exponential
suppression  of  fluctuations;  this  effect is  responsible  for  the
smallness  of  the  temperature  anisotropies.   However,  the  severe
experimental bounds  on violations of the  equivalence principle drive
some of the parameters in the model towards less natural numbers.

In our scenario, the end of inflation is triggered by the departure of
the self-interaction $W$ from its functional form during inflation. If
$W$  develops a  minimum about  which the  radion may  oscillate, then
reheating ensues. However, for  negatively curved extra dimensions the
minimum  of the potential  occurs at  a negative  value of  the energy
density, whence it is  problematic, though certainly not impossible to
reheat the universe.  This  mechanism, which ends inflation and causes
reheating,  also prevents  the radion  from violating  the equivalence
principle. Even though the potential of the radion-inflaton system has
a flat  direction, through an  appropriate choice of parameters  it is
possible to  project the  radion onto the  massive eigenvector  of the
mass-matrix.   Thus,   we  suppress  violations   of  the  equivalence
principle which  arise from varied coupling  of the radion  to matter. 
This idea  might be  of use,  in contexts different  from ours,  as an
alternative method  for decoupling the radion from  matter in theories
with flat directions.

\begin{acknowledgments}
  We  thank  Robert  Brandenberger  and Geraldine  Sevant  for  useful
  suggestions and valuable discussions.  CAP was supported by the U.S.
  DoE grant DE-FG02-90ER40560.
\end{acknowledgments}

\end{document}